\pdfoutput=1
\documentclass[sigconf]{acmart}

\AtBeginDocument{%
  }

\copyrightyear{2025}
\acmYear{2025}
\setcopyright{cc}
\setcctype{by-nc-sa}
\acmConference[CHI '25]{CHI Conference on Human Factors in Computing
Systems}{April 26-May 1, 2025}{Yokohama, Japan}
\acmBooktitle{CHI Conference on Human Factors in Computing Systems (CHI
'25), April 26-May 1, 2025, Yokohama,
Japan}\acmDOI{10.1145/3706598.3713370}
\acmISBN{979-8-4007-1394-1/25/04}

\usepackage{soul}  
\usepackage{xspace}
\usepackage{color}
\usepackage{listings}
\usepackage{multirow}
\usepackage{makecell}
\usepackage[ruled,linesnumbered]{algorithm2e}

\usepackage{physics}
\usepackage{colortbl}

\newcommand{\ie}{{i.e.,}\xspace}
\newcommand{\eg}{{e.g.,}\xspace}
\newcommand{\cf}{{c.f.}\xspace}
\newcommand{\ea}{{et~al.}\xspace}

\newcommand{\etc}{{etc.}\xspace}

\newcommand{\bpstart}[1]{\vspace{1mm} \noindent{\textbf{#1.}}}

\newcommand{\bstartnc}[1]{\vspace{1mm} \noindent{\textbf{#1}}}

\begin{document}

\title{Toward Human-Quantum Computer Interaction: 
Interface Techniques for Usable Quantum Computing}

\author{Hyeok Kim}
\email{hyeokk@uw.edu}
\orcid{0000-0003-4340-4470}
\affiliation{%
  \institution{University of Washington}
  \city{Seattle}
  \state{Washington}
  \country{U.S.A.}
}

\author{Mingyoung J. Jeng}
\email{mingyoungjeng@u.northwestern.edu}
\orcid{0009-0007-4452-3435}
\affiliation{%
  \institution{Northwestern University}
  \city{Evanston}
  \state{Illinois}
  \country{U.S.A.}
}

\author{Kaitlin N. Smith}
\email{kns@northwestern.edu}
\orcid{0000-0002-1169-3696}
\affiliation{%
  \institution{Northwestern University}
  \city{Evanston}
  \state{Illinois}
  \country{U.S.A.}
}



\begin{abstract}
By leveraging quantum-mechanical properties like superposition, entanglement, and interference, quantum computing (QC) offers promising solutions for problems that classical computing has not been able to solve efficiently, such as drug discovery, cryptography, and physical simulation.
Unfortunately, adopting QC remains difficult for potential users like QC beginners and application-specific domain experts, due to limited theoretical and practical knowledge, the lack of integrated interface-wise support, and poor documentation.
For example, to use quantum computers, one has to convert conceptual logic into low-level codes, analyze quantum program results, and share programs and results. 
To support the wider adoption of QC, we, as designers and QC experts, propose interaction techniques for QC through design iterations.
These techniques include writing quantum codes conceptually, comparing initial quantum programs with optimized programs, sharing quantum program results, and exploring quantum machines.
We demonstrate the feasibility and utility of these techniques via use cases with high-fidelity prototypes.
\end{abstract}





\maketitle

\section{Introduction}\label{sec:intro}

Quantum computing (QC) offers promising solutions for classically intractable problems that classical computers (those with digital bits) have not been able to solve due to time complexity or memory limitations~\cite{Arute2019:quantumSupremacy}. 
For example, factoring a number into two numbers is useful for cryptography, yet classical algorithms typically take exponential time to solve it (days or years) whereas Shor's algorithm~\cite{shors:algorithm} for QC takes only logarithmic time (minutes or days, respectively). 
Quantum computers are particularly useful in solving problems that requires searching a few valid instances from tons of potential cases, such as drug discovery~\cite{cao2018:drug}, physics simulation~\cite{somma2003:physics}, and cryptography~\cite{shors:algorithm}.
With more advances in hardware and software architecture, QC is expected to empower various scientific and engineering areas. 

However, using quantum computers requires in-depth knowledge and skills, making it challenging for a wide range of domain experts to adopt QC as a feasible solution.
With current tooling like Qiskit~\cite{qiskit} or Strawberry Fields~\cite{strawberryfields}, developers need to write programs by specifying operations for qubits (where a qubit has a value of $|{\psi}\rangle =\alpha |{0}\rangle +\beta |{1}\rangle$) and read results in terms of bit strings rather than easily-interpretable formats like natural numbers or text.
Expert users often need to look at documentation to retrieve detailed information (\eg~physical properties of a machine), yet they are often too basic (\eg~``\texttt{qubit\_properties}: Return \texttt{QubitProperties} for a given qubit.''~\footnote{\url{https://docs.quantum.ibm.com/api/qiskit/qiskit.providers.BackendV2\#qubit_properties}}) or mathematically oriented, lacking example use cases.
QC platforms like IBM and AWS offer graphical tools to explore machines and programs, yet using them requires frequent switching between coding tools and them, and those interfaces often fail to show information with large programs. 

Therefore, more intuitive interfaces are highly necessary to mitigate low-level technical hurdles and hence foster QC to users in various domain areas.
In response, we, as a team of a design researcher and QC researchers, identify a set of design principles through a survey of existing QC tools and an iterative design process of building QC interfaces.
These principles include linking high-level conceptual ideas with low-level quantum information, providing support at different levels of computing, and applying usability standards.
Based on these principles, we showcase a set of interaction techniques for usable QC, such as conceptual program writer, hardware dashboard, program viewer, job-sharing API, and problem-specific result viewer.
Implementing these techniques as open-sourced high-fidelity prototypes, we demonstrate their usability and technical soundness through use cases that reflect QC learning and research.

\section{Background: Hello \textit{Quantum} World}\label{sec:bg}

We first describe quantum computing (QC) basics, practices, and tools that are relevant to understand this work. 
We note that the below overview includes simplification for surface-level understanding; refer to Nielsen and Chuang~\cite{mike:and:ike} for more precise descriptions.
Readers could also use this section as a glossary.

\subsection{Quantum Computing in a Nut Shell}\label{sec:bg:nutshell}

\begin{figure}
    \centering
    \includegraphics[width=\linewidth]{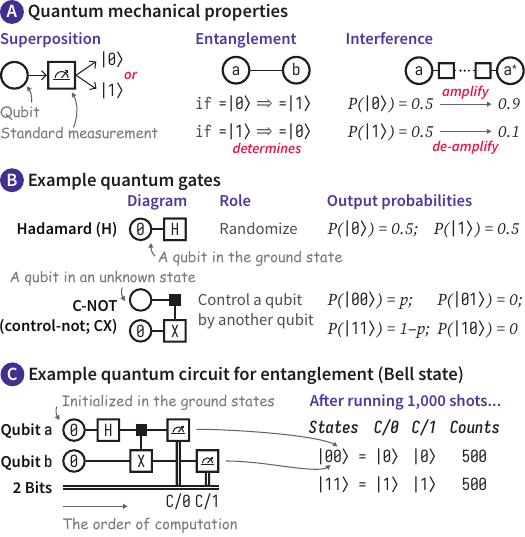}
    \caption{(A) Diagrams for quantum mechanical properties. (B) Example quantum gates. (C) An example quantum circuit that generates a Bell state or quantum entanglement using Hadamard and C-NOT gates on two qubits. Bra ($\langle{}|$) and ket ($|{}\rangle$) are a notation for representing quantum states.}
    \label{fig:qc-overview}
    \Description{There are three parts. A, quantum mechanical properties, superposition, entanglement, and interference. B, example quantum gates: Hadamard, Rotation, and C-Not gates and their output probabilities. C, an example quantum circuit for entanglement or bell state, with a Hadamard gate on qubit A, and a C-Not gate on qubits A and B, resulting in state 500 zero-zero state and 500 one-one state.}
\end{figure}

\noindent Quantum computers are defined as computing devices that utilize quantum mechanical properties: \textit{superposition}, \textit{entanglement}, and \textit{interference} of quantum particles, as depicted in \autoref{fig:qc-overview}A.
The base computation unit of quantum computers is called a \textbf{qubit}, standing for quantum bit, which are manipulated to exhibit those properties.
Bits of classical computers (those that we use everyday) are called classical bits or just bits. 
\textbf{Superposition} describes how quantum \textbf{state}s can exist in a probabilistic distribution of discrete measurement outcomes~\cite{mike:and:ike}. 
A measurement device measures a qubit by forcing its state into a fixed value from a few discrete values with corresponding probabilities (\ie~non-deterministic).
Existing quantum computers typically employ \textbf{standard measurement} that measures a qubit in either $|{0}\rangle$ (the ground state) or $|{1}\rangle$ (an excited state).
Superposition enables the generation of lots of candidate inputs for a given problem, allowing for inherent parallelism during computation.
Due to the probabilistic nature of QC, states are often prepared and measured over many trials to form a measurement distribution that represents a quantum application's outcome.

Next, \textbf{entanglement} describes how two pieces of quantum information must be considered as a composite system after a series of special interactions. 
Even if two entangled qubits are separated by a great distance, 
we can infer about the measurement of one by measuring the other~\cite{mike:and:ike}.
For example, suppose two qubits are entangled to have the same state.
If a qubit is measured as $|{0}\rangle$ (or $|{1}\rangle$), then its entangled counterpart is always measured as $|{0}\rangle$ (or $|{1}\rangle$).
Entanglement allows for accessing and manipulating information using its correlation with other information. 
Lastly, \textbf{interference} means that a manipulation on a qubit affects or cancels previous manipulations on the same qubit.
Interference is useful to emphasize desired outcomes or de-emphasize unwanted outcomes. 

A quantum computer employs a physical processor to apply superposition, entanglement, and interference to a qubit(s).
A high-level classification of quantum computers is whether they manipulate qubits by continuously changing their parameters (\textbf{analogue}) or using discrete gates (\textbf{digital} or \textbf{gate-based}).
In this work, we primarily focus on digital quantum computers given its wide availability in practice (\eg~IBM, AWS). 
For example, a \textbf{Hadamard gate} (or \textit{H gate}) sets a qubit to be measured as 0 or 1 with a probability of 50\%, each, implementing the superposition of qubits (\autoref{fig:qc-overview}B).
By taking two qubits, a \textit{Control-NOT} (or \textbf{C-NOT}, \textit{CX}) \textit{gate} flips the state of one qubit ($|{0}\rangle \rightarrow |{1}\rangle$, vice versa) only if the other qubit's state is $|{1}\rangle$. 
The state resulting from these operations on two qubits is also called \textbf{Bell state}.
An H gate on a qubit and a C-NOT gate on it and another qubit enables the entanglement between them (\autoref{fig:qc-overview}C).
By applying multiple consecutive gates on the same qubits, one can realize interference. 

A quantum program is typically called a \textbf{(quantum) circuit}, composed of qubits and gates on those qubits.
QC developers must be able to deal with this abstraction if they want to debug or create QC algorithms.
A quantum circuit is usually represented as in \autoref{fig:qc-overview}C, where each horizontal line represents a qubit and boxes and dots on those lines represent gates applied to the corresponding qubits. 
The gates in each column in the diagram are those that can be operated at the same time (\ie~in the same \textbf{layer}).
After computing all the specified gates, selected qubits are measured as either $|{0}\rangle$ or $|{1}\rangle$. 
The measured outcomes are recorded on corresponding classical bits as indicated using the horizontal double-line at the bottom in the circuit diagram.
By running the same quantum circuit lots of times, or \textbf{shots}, (typically $1,000$+ shots), one can obtain the probability distribution of outcome states. 
For example, given a circuit where two qubits $a$ and $b$ are manipulated to a Bell state, the measured outcomes after $1,000$ shots should theoretically have $500$ $|{00}\rangle$ states and $500$ $|{11}\rangle$ states.

\subsection{Quantum Computing Practices}\label{sec:bg:practice}
\noindent To build and run a quantum program or circuit, a developer typically needs to perform the following tasks: writing a quantum program, selecting a machine(s) to use, optimizing a circuit for the machine to use, running the circuit, and analyzing the output of the program. 
To help understand, one can make an analogy to data science procedures that commonly include implementing initial statistical or learning models, revising them with different parameters, finding a remote machine if complex computation is needed, and analyzing the computation results like model fits, uncertainty, accuracy, and precision~\cite{crisan2021}. 
Below, we overview QC practices based on official tutorials of QC platforms~\cite{qiskitGuide,cirqGuide,braketGuide}.

First, a developer must express the problem they want to solve using qubits and gates. 
This is challenging because quantum gates are defined mathematically, and hence their relationships to the conceptual ideas about a problem (\eg~factoring, desired chemical properties) are not necessarily obvious. 
Thus, finding a quantum algorithm for a real-world problem is an important research topic in QC.
For instance, Quantum Algorithm Zoo\footnote{\url{https://quantumalgorithmzoo.org/}} has collected problems and algorithms for QC, and PennyLane Datasets\footnote{\url{https://pennylane.ai/datasets/}} offers a collection of data to use in QC programs.

Before running the quantum circuit on an actual QC machine (quantum processing unit, \textbf{QPU}), the developer needs to decide which machine to use by considering several factors, such as cost, availability (\eg~shut down, too many pending jobs), architecture, and error constraints.
For example, running a QPU on cloud for five minutes can cost a few hundred US dollars, developers need to select simulators first and then use real machines later.
Each machine exhibits different \textbf{gate error} rates (\ie~how likely a gate is to cause an error) and \textbf{decoherence time} (\ie~how long a qubit maintains its gate-manipulated state).
Given that physical properties must be assessed frequently due to their tenancy to vary over time~\cite{dasgupta2021stability}, the developer needs to check them before choosing a machine.

After deciding which device to use, the developer needs to \textbf{optimize} or \textbf{transpile} the initial circuit (\textbf{logical circuit}) into a much larger \textbf{physical circuit}. 
For example, a simple 10-gate logical circuit can result in a 500-gate physical circuit.
Optimization is necessary because a physical machine only supports a specific set of \textbf{basis gates} which is smaller than that of gates in logical circuits.
This is similar to decomposing a wide range of classical logic gates using a smaller set of gates (\eg~an IF classical gate decomposed using NOT and OR classical gates).
By merging and decomposing logical gates, optimization also aims to minimize the cumulative gate error rate by reducing the total physical gates and selecting least erroneous qubits and gates.
Given exponential possibilities to optimize a logical circuit, many optimization algorithms rely on random seeds and other parameters.
We note that this procedure is algorithmic optimization while device-level optimization directly deals with controlling physical devices.

After running a transpiled circuit on a physical machine, the developer must carefully analyze the measurement output. 
Current, \textbf{noisy intermediate-scale quantum (NISQ) era} machines exhibit non-negligible error rates as hinted earlier, the final output could include some erroneous computations.
For example, after running a simple entanglement circuit shown in \autoref{fig:qc-overview}C for 1,000 shots, the theoretical output must include 500 $|{00}\rangle$ and 500 $|{11}\rangle$ only. 
However, an actual output may include several $|{01}\rangle$ or $|{10}\rangle$, and $|{00}\rangle$ and $|{11}\rangle$ tend to be measured different times. 

\begin{table*}[t]
    \centering
    \caption{An overview of QC tools' interface-wise support for QC tasks in comparison to our prototypes (present work). For N/A, QuTip does not have associated machines and qBraid offers an IDE for using other quantum programming toolkits like Qiskit or PennyLane. Classiq's library offers declarative scripting as well as procedural programming (hybrid). Under the present work, `(Qiskit)' means that we use relevant features from Qiskit.}
    \label{tab:qc-tools}
\setlength{\tabcolsep}{3pt}
\def\arraystretch{1.18}%
\resizebox{\textwidth}{!}{\sffamily
\begin{tabular}{l|l|l|l|l|l|l|l|l|l|l|l}
{\color[HTML]{6200C9} \textbf{Tools}}                                            & \cellcolor[HTML]{FFF7F1}{\color[HTML]{6200C9} \textbf{Present work}} & {\color[HTML]{6200C9} \textbf{Qiskit}} & {\color[HTML]{6200C9} \textbf{Cirq}} & {\color[HTML]{6200C9} \textbf{QuTip}}                          & {\color[HTML]{6200C9} \textbf{StrawberryFields}} & {\color[HTML]{6200C9} \textbf{PennyLane}} & {\color[HTML]{6200C9} \textbf{Braket}} & {\color[HTML]{6200C9} \textbf{CUDA-Q}} & {\color[HTML]{6200C9} \textbf{Azure}} & {\color[HTML]{6200C9} \textbf{qBraid}} & {\color[HTML]{6200C9} \textbf{Classiq}} \\
\rowcolor[HTML]{EFEFEF} 
Developed by                                                                     & \cellcolor[HTML]{EEE6E0}-                                            & IBM                                    & Google                               & QuTip                                                          & Xanadu                                           & Xanadu                                    & AWS                                    & NVDIA                                  & Microsoft                             & qBraid                                 & Classiq                                 \\
\textit{\textbf{Circuit composition}}                                            & \cellcolor[HTML]{FFF7F1}                                             &                                        &                                      &                                                                &                                                  &                                           &                                        &                                        &                                       &                                        &                                         \\
\rowcolor[HTML]{EFEFEF} 
Programming-based                                                                & \cellcolor[HTML]{EEE6E0}Hybrid                                       & Declarative                            & Declarative                          & Declarative                                                    & Declarative                                      & Declarative                               & Declarative                            & Declarative                            & Entry-point                           & Use others                             & Hybrid                                  \\
Graphical interface                                                              & \cellcolor[HTML]{FFF7F1}Yes                                          & {\color[HTML]{9B9B9B} No}              & {\color[HTML]{9B9B9B} No}            & {\color[HTML]{9B9B9B} No}                                      & {\color[HTML]{9B9B9B} No}                        & {\color[HTML]{9B9B9B} No}                 & {\color[HTML]{9B9B9B} No}              & {\color[HTML]{9B9B9B} No}              & {\color[HTML]{9B9B9B} No}             & Composer                               & Jamboard                                \\
\rowcolor[HTML]{EFEFEF} 
Predefined subroutines                                                           & \cellcolor[HTML]{EEE6E0}Yes                                          & {\color[HTML]{9B9B9B} No}              & {\color[HTML]{9B9B9B} No}            & {\color[HTML]{9B9B9B} No}                                      & {\color[HTML]{9B9B9B} No}                        & Templates                                 & {\color[HTML]{9B9B9B} No}              & {\color[HTML]{9B9B9B} No}              & {\color[HTML]{9B9B9B} No}             & {\color[HTML]{C0C0C0} N/A}             & Templates                               \\
Defining a custom gate                                                           & \cellcolor[HTML]{FFF7F1}(Qiskit)                                     & Yes                                    & Yes                                  & Yes                                                            & Yes                                              & Yes                                       & Yes                                    & Yes                                    & Yes                                   & {\color[HTML]{C0C0C0} N/A}             & Yes                                     \\
\rowcolor[HTML]{EFEFEF} 
{\color[HTML]{000000} Conceptual operations}                                     & \cellcolor[HTML]{EEE6E0}{\color[HTML]{000000} Yes}                   & {\color[HTML]{9B9B9B} No}              & {\color[HTML]{9B9B9B} No}            & {\color[HTML]{9B9B9B} No}                                      & {\color[HTML]{9B9B9B} No}                        & {\color[HTML]{000000} Yes}                & {\color[HTML]{9B9B9B} No}              & {\color[HTML]{9B9B9B} No}              & {\color[HTML]{9B9B9B} No}             & {\color[HTML]{C0C0C0} N/A}             & {\color[HTML]{000000} Yes}              \\
Circuit visualization                                                            & \cellcolor[HTML]{FFF7F1}Interactive                                  & Traditional                            & Traditional                          & Traditional                                                    & Traditional                                      & Traditional                               & Traditional                            & Traditional                            & Traditional                           & {\color[HTML]{C0C0C0} N/A}             & Diagrams                                \\
\rowcolor[HTML]{EFEFEF} 
Qubit state visualization                                                        & \cellcolor[HTML]{EEE6E0}{\color[HTML]{C0C0C0} No}                    & Bloch                                  & Bloch                                & \begin{tabular}[c]{@{}l@{}}Bloch, \\ energy level\end{tabular} & {\color[HTML]{9B9B9B} No}                        & {\color[HTML]{9B9B9B} No}                 & {\color[HTML]{9B9B9B} No}              & Bloch                                  & Bloch                                 & {\color[HTML]{C0C0C0} N/A}             & {\color[HTML]{9B9B9B} No}               \\
Animated visualization                                                           & \cellcolor[HTML]{FFF7F1}Yes                                          & {\color[HTML]{9B9B9B} No}              & {\color[HTML]{9B9B9B} No}            & Qubits                                                         & {\color[HTML]{9B9B9B} No}                        & {\color[HTML]{9B9B9B} No}                 & {\color[HTML]{9B9B9B} No}              & {\color[HTML]{9B9B9B} No}              & {\color[HTML]{9B9B9B} No}             & {\color[HTML]{C0C0C0} N/A}             & {\color[HTML]{9B9B9B} No}               \\
\rowcolor[HTML]{EFEFEF} 
AI support                                                                       & \cellcolor[HTML]{EEE6E0}{\color[HTML]{9B9B9B} No}                    & {\color[HTML]{9B9B9B} No}              & {\color[HTML]{9B9B9B} No}            & {\color[HTML]{9B9B9B} No}                                      & {\color[HTML]{9B9B9B} No}                        & {\color[HTML]{9B9B9B} No}                 & {\color[HTML]{9B9B9B} No}              & {\color[HTML]{9B9B9B} No}              & Yes                                   & {\color[HTML]{9B9B9B} No}              & {\color[HTML]{9B9B9B} No}               \\
\textit{\textbf{Machine selection}}                                              & \cellcolor[HTML]{FFF7F1}                                             &                                        &                                      &                                                                &                                                  &                                           &                                        &                                        &                                       &                                        &                                         \\
\rowcolor[HTML]{EFEFEF} 
Filtering a machine                                                              & \cellcolor[HTML]{EEE6E0}(Qiskit)                                     & Advanced                               & {\color[HTML]{9B9B9B} No}            & {\color[HTML]{C0C0C0} N/A}                                     & Basic                                            & Basic                                     & Basic                                  & Basic                                  & Basic                                 & Basic                                  & Basic                                   \\
Status dashboard                                                                 & \cellcolor[HTML]{FFF7F1}On IDE                                       & On platform                            & {\color[HTML]{9B9B9B} No}            & {\color[HTML]{C0C0C0} N/A}                                     & On platform                                      & On platform                               & On platform                            & On platform                            & On platform                           & On platform \& IDE                     & On IDE                                  \\
\rowcolor[HTML]{EFEFEF} 
Status visualization                                                             & \cellcolor[HTML]{EEE6E0}On IDE                                       & Template                               & Tempalte                             & {\color[HTML]{C0C0C0} N/A}                                     & {\color[HTML]{9B9B9B} No}                        & {\color[HTML]{9B9B9B} No}                 & {\color[HTML]{9B9B9B} No}              & {\color[HTML]{9B9B9B} No}              & {\color[HTML]{9B9B9B} No}             & {\color[HTML]{9B9B9B} No}              & {\color[HTML]{9B9B9B} No}               \\
\textit{\textbf{Optimization}}                                                   & \cellcolor[HTML]{FFF7F1}                                             &                                        &                                      & {\color[HTML]{C0C0C0} }                                        &                                                  &                                           &                                        &                                        &                                       &                                        &                                         \\
\rowcolor[HTML]{EFEFEF} 
Parameterizing optimizers                                                        & \cellcolor[HTML]{EEE6E0}(Qiskit)                                     & Yes                                    & Yes                                  & {\color[HTML]{C0C0C0} N/A}                                     & Yes                                              & Yes                                       & {\color[HTML]{9B9B9B} No}              & {\color[HTML]{9B9B9B} No}              & {\color[HTML]{9B9B9B} No}             & {\color[HTML]{C0C0C0} N/A}             & Yes                                     \\
Reviewing optimization                                                           & \cellcolor[HTML]{FFF7F1}Interactive                                  & Yes                                    & Yes                                  & {\color[HTML]{C0C0C0} N/A}                                     & Yes                                              & Yes                                       & Yes                                    & {\color[HTML]{9B9B9B} No}              & {\color[HTML]{9B9B9B} No}             & {\color[HTML]{C0C0C0} N/A}             & {\color[HTML]{9B9B9B} No}               \\
\rowcolor[HTML]{EFEFEF} 
Comparing optimizations                                                          & \cellcolor[HTML]{EEE6E0}Yes                                          & {\color[HTML]{9B9B9B} No}              & {\color[HTML]{9B9B9B} No}            & {\color[HTML]{C0C0C0} N/A}                                     & {\color[HTML]{9B9B9B} No}                        & {\color[HTML]{9B9B9B} No}                 & {\color[HTML]{9B9B9B} No}              & {\color[HTML]{9B9B9B} No}              & {\color[HTML]{9B9B9B} No}             & {\color[HTML]{9B9B9B} No}              & {\color[HTML]{9B9B9B} No}               \\
\textit{\textbf{Result analysis}}                                                & \cellcolor[HTML]{FFF7F1}                                             &                                        &                                      &                                                                &                                                  &                                           &                                        &                                        &                                       &                                        &                                         \\
\rowcolor[HTML]{EFEFEF} 
Measurement histogram                                                            & \cellcolor[HTML]{EEE6E0}On IDE                                       & Function                               & Function                             & Function                                                       & Templates                                        & Templates                                 & No                                     & No                                     & On IDE                                & On IDE                                 & On IDE                                  \\
\begin{tabular}[c]{@{}l@{}}Problem-specific\\ result representation\end{tabular} & \cellcolor[HTML]{FFF7F1}Yes                                          & {\color[HTML]{9B9B9B} No}              & {\color[HTML]{9B9B9B} No}            & {\color[HTML]{9B9B9B} No}                                      & Templates                                        & Templates                                 & {\color[HTML]{9B9B9B} No}              & {\color[HTML]{9B9B9B} No}              & {\color[HTML]{9B9B9B} No}             & {\color[HTML]{9B9B9B} No}              & {\color[HTML]{9B9B9B} No}               \\
\rowcolor[HTML]{EFEFEF} 
Uncertainty                                                                      & \cellcolor[HTML]{EEE6E0}Yes                                          & {\color[HTML]{9B9B9B} No}              & {\color[HTML]{9B9B9B} No}            & {\color[HTML]{9B9B9B} No}                                      & {\color[HTML]{9B9B9B} No}                        & {\color[HTML]{9B9B9B} No}                 & {\color[HTML]{9B9B9B} No}              & {\color[HTML]{9B9B9B} No}              & {\color[HTML]{9B9B9B} No}             & {\color[HTML]{9B9B9B} No}              & {\color[HTML]{9B9B9B} No}              
\end{tabular}
}
    \Description{A summary table for the present work and quantum computing tools: qiskit, cirq, qu-tip, strawberry fields, penny lane, braket, cuda-q, zaure, q-braid, classi-q.}
\end{table*}

\subsection{A Survey of Quantum Computing Tools}\label{sec:bg:tools}

To provide a snapshot of the state-of-the-art QC tools, we surveyed ten open-sourced libraries for gate-based QC in terms of user interfaces for supporting the above tasks. 
Specifically, we looked into Qiskit by IBM~\cite{qiskit}, Cirq by Google~\cite{cirq}, QuTip~\cite{qutip}, StrawberryFields~\cite{strawberryfields} and PennyLane~\cite{pennylane} by Xanadu, Braket by AWS~\cite{braket}, CUDA-Q by NVIDIA~\cite{cudaq}, Azure by Microsoft~\cite{azureQuantum}, qBraid~\cite{qbraid}, and Classiq~\cite{classiq}.
Our survey is summarized in \autoref{tab:qc-tools}.

\begin{figure}
    \centering
    \includegraphics[width=\linewidth]{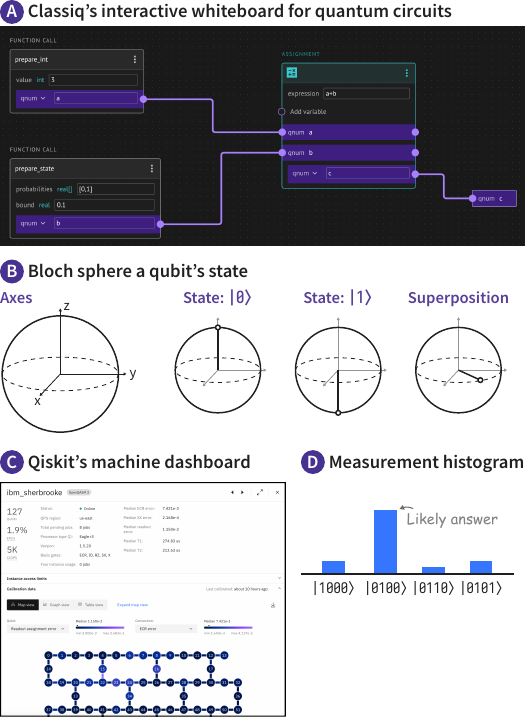}
    \caption{Visualizations offered by QC tools. (A) The interactive whiteboard of Classiq~\cite{classiq} (B) Bloch sphere. (C) Qiskit's machine status dashboard~\cite{qiskit} (D) Measurement histogram.}
    \label{fig:qc-vis}
    \Description{Four visualization cases for quantum computing. A, Classi-q's interactive whiteboard for quantum circuits. B, Bloch spheres for a qubit's state. A block sphere has three axes: x, y, and z. C, qiskit's machine dashboard. D, measurement histogram.}
\end{figure}

\bpstart{Circuit composition}
Most QC tools offer a Python library for specifying a circuit in a declarative way. 
Commonly, a developer first need to define a baseline logical circuit in terms of the number of qubits (for operations) and classical bits (for storing measurement outcomes) and then add gates to the circuit.
Using Cirq~\cite{cirq}, for example, a Bell state circuit can be written as \texttt{circuit.H(0).CNOT(0, 1)}, where \texttt{H} and \texttt{CNOT} stands for Hadamard and C-NOT gates, and \texttt{0} and \texttt{1} are the indices for the first two qubits.
Azure Quantum~\cite{azureQuantum} provides its own programming language, Q\#, which needs to define an entry point, similar to C\# and JAVA. 
qBraid~\cite{qbraid} and Classiq~\cite{classiq} provide graphical interfaces in addition to scripting.
qBraid offers a circuit composer~\cite{quirk} where a developer can drag and drop gates on a circuit diagram. 
Since qBraid's Python library is a wrapper for others like Qiskit and PennyLane, so developers could use expressions from their preferred libraries.
Classiq provides an interactive whiteboard (\eg~FigJam, Jamboard), where one can add a post-it for operations and connect them with lines (\autoref{fig:qc-vis}A).

PennyLane and Classiq provide templates for predefined subroutines (\eg~search algorithms, chemical processes) and conceptual representations (\eg~adding numbers, atom representations), reducing the gap between high-level program logic and low-level quantum circuits in part. 
Yet, developers need to manually change the code for their own use cases.
Most tools allow for defining a custom gate by passing a \textbf{unitary matrix} (of which the inverse and conjugate transposes are same). 
For visual representations, most tools provide internal functions to draw a traditional circuit visualization (\eg~\autoref{fig:qc-overview}C).
A few tools support drawing a \textbf{Bloch sphere} (\autoref{fig:qc-vis}B) that show a single qubit's expected state on a sphere.
QuTip supports creating animations for how qubits' states evolve over operations.
Azure supports generative AI-based assistance (Copilot~\cite{copilot}) for writing codes for quantum circuits.

\bpstart{Machine selection}
In many cases, QC tools let developers filter a machine by its name and provider (the company that owns the machine). 
Qiskit offers more advanced filters like the number of qubits in a machine, availability, the maximum number of shots, \etc
Many QC tools provide separate websites for a dashboard about machines' status (\eg~\autoref{fig:qc-vis}C).
Such dashboards are not directly accessible on an integrated development environment (IDE) like Jupyter Notebook~\cite{jupyterNotebooks}.
While qBraid and Classiq integrate machine status information on their IDE's, the level of available information is shallow (\eg~not showing error rates). 
Qiskit and Cirq's documentations provide templates for how to visualize qubit and operation errors on IDE.

\bpstart{Optimization}
Qiskit~\cite{qiskit}, Cirq~\cite{cirq}, StrawberryFields~\cite{strawberryfields}, PennyLane~\cite{pennylane}, and Classiq~\cite{classiq} let developers customize an optimizer by providing parameters like a random seed, the degree of optimization, and the use of different algorithms. 
As optimization procedures typically involve randomization due to exponential possibilities, setting different parameters can result in varying optimization outcomes for the same circuit, which can impact the quality of the program output.
Given that optimized circuits share the same data structure with logical circuits, developers usually visualize them for review.
Yet, CUDA-Q~\cite{cudaq}, Braket~\cite{braket}, and Azure~\cite{azureQuantum} incorporated optimization with execution functions, and Classiq~\cite{classiq} does not provide optimization results. 
However, none of these tools provide interfaces for comparing different optimization results; instead, a developer must come up with their own approach. 

\bpstart{Results analysis}
Most tools offer ways to visualize measurement outcomes (counts) as a histogram (\autoref{fig:qc-vis}D), such as a simple function, code templates as tutorials, and small panels in an IDE.
Given that quantum results take a form of bit string (0101...), translating them for problems (\eg~integer, letter, \etc) is a necessary step to interpret measurement outcomes.
StrawberryFields and PennyLane's tutorials offer several ways to do so, such as decision boundaries\footnote{\url{https://pennylane.ai/qml/demos/tutorial_kernels_module/}}. 
While measurement outcomes involve some errors, related uncertainty information is not usually communicated. 

\bpstart{Takeaways}
We summarize a few takeaways from this survey with respect to common usability support.
(1) \textit{QC libraries provide raw visualization and circuit codes as templates} in their tutorials rather than application programming interfaces (APIs) integrated to themselves. 
Because QC users are not necessarily classical programming experts and it is challenging to edit raw codes~\cite{mei2018:vistoolsurvey,battle2022:d3}, more integrated interface support for frequently used visualizations seems to be useful.
(2) \textit{Useful information is often not accessible on IDE.}
QC developers often need to switch platform's dashboard, IDE, documentations, and tutorials, which can make their tasks more tedious than needed.
(3) \textit{It is hard to make interactions for on-demand details.}
Visualization templates and APIs are often designed in a static way, limited to support chained interactions for seeking details.
For example, static circuit visualization for large circuits can be confusing to identify the relationship between qubits and gates.
To fill this gap, our work aims to find methods to integrate this kind of usability support for QC tools.

\section{Related Work}\label{sec:rw}

Our work is grounded by prior work on interface and visualization techniques for quantum computing and other relevant domains.

\subsection{Interfaces for Quantum Computing}\label{sec:rw:interfaces}

\noindent
Prior work in understanding how people use QC tools sets useful context for our work.
By interviewing seven scientists in different domains, Ashktorab~\ea~\cite{ashktorab2019:hqci} report the needs for assisting beginners with using QC tools with limited mathematical knowledge.
They also note that after passing a certain point in training, one might need and want to move on to more professional tools for a higher degree of freedom.
Furthermore, they recognize the importance of sharing and replicating quantum programs as well as visualizing quantum information in a scalable way given that $n$ qubits result in $2^n$ states (worst case). 
Prior work based on education settings~\cite{mykhailova2022:softwareQC,hu2024:qcLearning,tappert2019:experience,salehi2022:csQC} focused on the implementation of simpler circuits representing basic concepts, such as superposition and entanglement, and provided conceptual understanding of complex algorithms like quantum chemistry or optimization problems.
In addition, by analyzing self-paced interactive textbook, Wootton~\ea~\cite{wootton2021:interactiveTextbook} found that learners tended to stop at earlier chapters and frequently referred to appendix for linear algebra.
These findings indicate two competing goals for QC support: introducing complex quantum algorithms in a conceptual way vs. training learners to work with low-level components of those algorithms.
This work proposes and demonstrates approaches to conceptually representing components of quantum algorithms while not blocking the degree of freedom offered by low-level toolkits.

Visualization takes an important role in making QC more understandable~\cite{bethel2023:qcvis}, such as Bloch sphere~\cite{bloch1946} for qubit states, circuit diagrams, and measurement histograms for QC output.
On top of those common visual representations, prior approaches focused on interactive interfaces for circuit composition and result analysis.
First, interactive circuit composers like IBM Quantum Composer~\cite{ibmComposer} and Quirk~\cite{quirk} let users drag and drop gates on a circuit visualization.
Yet, those tools still require a deeper understanding of how each gate functions with respect to bigger conceptual ideas, which Classiq~\cite{classiq} is trying to address with an interactive whiteboard (\autoref{fig:qc-vis}A).
In addition, to improve the interpretability of changes to qubits' states along with operations in a circuit, prior work proposed approaches to integrating this information on a circuit diagram.
For instance, QuFlow~\cite{lin2018:quflow} shows possible states of qubits with their probabilities in number on a circuit, and QuantumEyes~\cite{ruan2024:quantumeyes} shows them as an area chart with small diagrams explaining how each gate affects the probabilities of qubit states. 
VENUS~\cite{ruan2023:venus} adopts geometrical probability representations for multi-qubit states given that traditional 3D representations like Bloch spheres are limited to show multi-qubit states effectively. 
However, showing qubit state probabilities on a circuit are feasible only when a circuit includes a few qubits because probability information is not tractable for circuits with more than a few qubits, which is why we need QC.
Instead, Quantivine~\cite{wen2024:quantivine} abstracts a circuit representation in terms of structural patterns (\eg~repeated gates) and allows for filtering a few qubits for more details like their interactions with other qubits.
Note that one can easily encode the structure of a circuit using Qiskit and check their decompositions.
Taking one step further, our work proposes and implements interactive visualizations for comparing logical and physical circuits and inspecting how physical qubits' status change over the execution of a circuit.

In analyzing QC results, developers need to consider non-negligible amount of gate and readout errors as well as adapt the histogram-based outputs for specific problem types.
To enable that, VACSEN~\cite{ruan2023:vacsen} parallelly shows quantum circuits with the decoherence time and error rates of the gates and measurements involved in executing the circuits.
In particular, VACSEN shows error information with timestamps because QC machines' physical properties change over time. 
To better support QC-based machine learning (quantum machine learning or QML), VIOLET~\cite{ruan2024:violet} offers a expert-targeted dashboard that shows QC outputs in the context of neural network (\eg~output states as features that a model has learned).
Yet, prior work does not offer a direct way to represent QC outputs with error rates and/or in a human-understandable format, which our work aims to propose approaches to complementing.

In adopting these complex computing approaches, developers frequently use computational notebooks (\eg~Jupyter Notebook) as they are easy to share, reaccess (\eg~compared to running them on a terminal), and refine~\cite{wang2024:supernova}.
Computational notebooks allow for integrating custom widgets, making it easier to combine work pipelines~\cite{wang2024:supernova,harden2024:hni}.
Yet, prior approaches in QC and complex computing methods tend to showcase standalone tools rather than those integrated with working environments like computational notebooks.
In addition, they often lack support for tasks pertaining to scripting and sharing codes for using them.
To address these integration-wise challenges, our work aims to propose interaction techniques for a wider range of QC lifecycle by integrating them within computational notebooks.

\subsection{Interface Support for Data-intensive Work}\label{sec:rw:others}

\noindent
Prior work has proposed interfaces for different tasks in data-intensive work pipelines, such as data analysis and statistical modeling.
Given data work is often done on a Jupyter notebook environment, for example, prior work suggested widgets for various tasks like developing statistical models~\cite{jun2022:tisane} and analyzing text data~\cite{lam2024:lloom}.
Dashboard-based interfaces are also common for complex statistical analysis~\cite{liu2021:boba,sarma2024:milliways} and literature analysis~\cite{narechania2022:vitality}.
Those interfaces commonly offer descriptions regarding steps that a user is taking. 
On the other hand, current QC users need to use the same tool regardless of their tasks because many tasks are constrained by QPUs provided via those tools.
For instance, gates defined using Cirq can be optimized only by Cirq because the optimization algorithms rely on QPUs accessed via Cirq.
Furthermore, users need to frequently switch between the above described QC tasks, requiring interfaces overarching those tasks.
Thus, our work also explores how to adapt prior user interface techniques across various QC tasks.
\section{Design Iterations}\label{sec:methods}
To derive design principles and interaction techniques for usable quantum computing interfaces, we went through an iterative design procedures by implementing working prototype interfaces.

\subsection{Methods Overview}
Our team consists of two QC researchers (Smith and Jeng) and an expert interface designer (Kim), showing varying degrees of expertise in QC.
Smith is a faculty member with several years of academic and industry QC research experience, and Jeng is a Ph.D. student specialized in quantum machine learning (QML) with several first-authored related papers.
Kim has a Ph.D. degree in HCI with baseline education in QC theory and practice by taking two graduate-level courses.

The team frequently (two to three times a week) discussed about prototype designs via remote meetings, Slack conversations, and GitHub issues.
As Kim led prototyping and maintained the code base on a GitHub repository, Smith and Jeng tested prototypes by applying them to their use cases and provided feedback via GitHub issues.
We had Zoom meetings to solicit higher level feedback and ideas like overall design directions and maintained meeting notes.
We shared useful resources like QC application cases and had offline discussion on a Slack channel.
This design iteration took place over about three months in total, where Jeng joined later in the process (for the last six weeks). 

\subsection{Identified Challenges}
Below, we describe challenges in QC that we have discussed to motivate our prototypes as well as those we found while implementing high-fidelity prototypes.

\bpstart{(C1) Difficulty in expressing ideas and interpreting results conceptually}
A recurring topic during our design iterations was the unmatched connection between conceptual ideas and quantum information.
In both theoretical and applied QC courses, for example, Kim learned Shor's algorithm~\cite{shors:algorithm} for factoring a large natural number into two other numbers, which is expected to advance quantum cryptography. 
However, he learned only about theoretical proofs and potential application areas (which is common in many QC classes~\cite{hu2024:qcLearning,tappert2019:experience}), but was not able to apply the algorithm to novel problems. 
It is tricky to try out different numbers for Shor's algorithm because doing so requires more profound knowledge than what typical beginners would have.
Qiskit provides a tutorial for how to do so\footnote{\url{https://github.com/qiskit-community/qiskit-community-tutorials/blob/master/algorithms/shor_algorithm.ipynb}}, but it only shows a case where one factor is fixed at 15. 
To apply a different factor, Kim had to come up with a method to encode that number using lots of X and Swap gates\footnote{A Swap gate changes the states of two qubits}, which was beyond Kim's initial capability (\cf~\cite{beauregard2003:shors}).

Similarly, when Jeng was developing image-convolution techniques for a QML project, she had to come up with a relatively large suite of custom codes to convert images to quantum states (and vice versa). 
Currently, QML developers lack shared standards that work generically with other image-processing algorithms.
Instead, they often need to edit codes from tutorials, making it tedious to apply those codes for novel use cases. 
Even though more experienced in QC research, Smith was not necessarily able to immediately come up with those methods because Jeng focused on specific QML techniques while Smith was more interested scalable optimization techniques across different application scenarios.

The same problem of unmatched levels of information persisted in interpreting outcomes of quantum programs.
QC outputs are typically expressed as a histogram of each state vector's count.
For Kim, it was difficult to interpret what each state vector means when there are more than a few of them. 
From the previous Shor's algorithm example, the main goal is to detect a periodic pattern from the output.
Common histogram codes from tutorials usually just show the counts as bar charts.
In doing so, they omit unmeasured state vectors and produce overly dense views, making it to difficult to quickly find patterns.

On the other hand, Jeng's above project needed to encode those state vector counts as images to see whether the QML procedures were successful. 
While each state vector count is often interpreted as the likelihood of being an answer, Jeng needed to convert the entire distribution as an image. 
She had to manually implement this conversion algorithm on her own.
The lack of well-packaged libraries caused uncertainty in debugging---whether classical parts or quantum parts were wrong. 

\bpstart{(C2) Diverse needs in navigating QC hardware space}
To best use NISQ era quantum machines, a developer must be aware of low-level technical properties to an extent.
However, it is often hard to do so because APIs often lack consistency in their design.
For example, Kim attended Smith's live demo of her work pipelines during her lecture before the design iterations. 
The class discussed how confusing it is to write codes for obtaining the properties of a qubit because of this inconsistency.
For instance, Qiskit offers \texttt{properties}, \texttt{configuration}, and \texttt{options} methods to retrieve data with slight differences in usage and available information; some information can be obtained via both \texttt{properties} and \texttt{configuration} while others are only available from either one of them.
This inconsistency does not necessarily indicate wrong design but mean that the machine providers intended those APIs to reflect hardware-level constraints (\eg~how they extract information from the machine) that individual programmers have limited controls on.

In addition to the difficulty in navigating the vast space of hard\-ware-level information, each task needed different types of physical properties.
For example, Smith posted an issue on GitHub to discuss the need for an ability to review different optimization parameters and outcomes. 
Smith suggested this feature because, as a QC architecture researcher, it was one of her common tasks to compare how different optimization parameters affect the fidelity (\eg~the comparison of the machine results to the ideal outcome) of the resulting physical circuits. 
Kim initially did not sense the need to compare optimization outcomes primarily because his focus was more on applications and algorithms.
For Kim, inspecting optimization outcome could be tedious as small (utility scale) application algorithms often result in 4-5000 physical gates.
Instead, Kim was more interested in pending jobs on a machine, overall fidelity, \etc

While optimization had not been a primary interest of Jeng, she recently started working on architecture-related research, so she suggested an ability to see \textit{pulse} information. 
The \textit{pulse} data of a QPU exhibit the frequencies and amplitudes of laser devices that the QPU uses to implement gates (\ie~manipulate qubits).
Jeng wanted to develop skills in pulse-level quantum programming that can further optimize circuits.
Although Smith mainly conducted research on QC optimization, her focus was on algorithmic methods, so she did not need support for device-level optimization.

\bpstart{(C3) Things working classically but not quantum-ly}
Widely-used practices among classical programmers are not necessarily directly transferrable to those of QC programmers with current environments.
For example, Smith and her collaborators often needed to share a circuit, run it on different machines and simulators, and compare the results in order to cross-validate algorithms and error-mitigation techniques.
To do so, Smith had two options: sharing a personal authentication token for IBM cloud or saving results in a single text file.
The first option is easier but creates obvious security issues.
Thus, Smith preferred the second option. 
However, the lack of standardization and portability bottlenecks of quantum programs and outcomes causes challenges associated with sharing data among researchers. 
This option requires detailed instructions for how to retrieve the transmitted quantum program data, which adds communication costs like back-and-forth chats.

While Kim was an experienced programmer, he has been familiar with dynamic memory allocation that changes the memory size of a variable as its value changes.
Modern programming languages like Python and JavaScript support dynamic memory allocation to make it easy to program. 
In contrast, QC programmers have to predefine the numbers of qubits and classical bits to use before writing the actual algorithm.
It is hard to define some offset qubits to allow for backend-based dynamic memory allocation because each qubit is a costly resource.
When Kim tried to make changes in the code to try new things, for example, he often forgot about making changes to qubits, resulting in bugs.
These challenges resonate the findings from our QC library survey.

\section{Design Principles for QC Interfaces}\label{sec:principles}

Based on our iterative design process and prior work in QC interfaces~\cite{ashktorab2019:hqci,lin2018:quflow,ruan2024:quantumeyes,ruan2023:venus,wen2024:quantivine,ruan2023:vacsen,ruan2024:violet}, we derive the following design principles to motivate our design decisions. 
Given the shared high-level objectives (QC interfaces), we use Ashktorab~\ea~\cite{ashktorab2019:hqci}'s interview study to compare with the above challenges along with other QC education literature~\cite{wootton2021:interactiveTextbook,mykhailova2022:softwareQC,hu2024:qcLearning,tappert2019:experience,salehi2022:csQC}.
We also considered prior work regarding data science work pipelines~\cite{wang2024:supernova,crisan2021} given their similarity in terms of high-level procedures and intensive use of computational notebooks.

\bpstart{(P1) Link conceptual ideas to quantum information}
As specialized for processing quantum mechanical properties, QC programs exhibit more tight association of their inputs and outputs with low-level quantum information, compared to classical computing~\cite{ashktorab2019:hqci}.
Studies also imply that an understanding of rudimentary concepts like qubits and gates does not necessarily mean an ability to apply and interpret actual circuits based on those concepts~\cite{mykhailova2022:softwareQC,wootton2021:interactiveTextbook,ashktorab2019:hqci}.
In our design iteration, for example, Jeng shared her experience in making custom classical programs to encode human-readable inputs like image files as quantum states for QML algorithms.
Similarly, when learning Shor's algorithm~\cite{shors:algorithm} that divides integers into two interger factors, Kim had struggles in understanding how the algorithm encodes integer multiplications as gates.
This generalization challenge makes it difficult to conceptually express QC programs, which in turn complicates learning real-world applications like cryptography.

On the other hand, QC program outputs are essentially the counts of measured qubit states that are expressed as bit strings (\eg~$|{00101011}\rangle$). 
For instance, Kim needed to manually convert the output bit strings of Shor's algorithm into human-readable numbers. 
Similarly, Jeng had to come up with custom functions to convert measurement outcomes (as probability distributions) into RGB images.
Furthermore, a QC program with $n$ qubits can result in up to $2^n$ states, indicating the need for scalability in expressing QC output. 
For example, the previous Shor's algorithm tutorial by Qiskit with a fixed factor of 15 requires measuring 8 qubits.
This measurement results in $2^8=256$ possible states on a real machine, where a histogram is not necessarily an ideal representation method due to the excessive visual density.
Ashktorab~\ea~\cite{ashktorab2019:hqci} also raise this scalability issue in quantum state and circuit visualizations because existing representation methods may fail to support interpreting circuits beyond a certain size.
Therefore, usable QC interfaces must support mapping developers' high-level ideas to low-level quantum information to improve flexibility in circuit composition and interpretability in result analysis. 

\bpstart{(P2) Provide support at different levels of computing}
Through the design iterations, we recognized that QC developers would work on varying levels of computing, like architecture, compiler, algorithms, \etc
For example, Smith wanted a lot of details in optimization outcomes while Kim did not.
Jeng's needs for optimization detail increased as she started working on architecture-related research.
Similarly, Ashktorab~\ea~\cite{ashktorab2019:hqci} identify different QC-related roles: \textit{scientists} who want full access to low-level details; \textit{developers} who do not necessarily want those in-depth information; and \textit{enthusiasts} who are interested in exploring QC capabilities with easy interactions. 
These different needs for information reflect the varying scales and goals in the practice and research of QC.

Flexible support at different scales is important for QC tooling because currently developers with varying scopes and goals have to use the same tools (\eg~Qiskit, CirQ).
For comparison, developers for classical computing use programming tools and libraries defined at different levels, such as JavaScript for user applications, Python and R for data science, C++ and Rust for compilers, assembly languages, \etc
In contrast, many QC toolkits are designed at a similar level of abstraction (\ie~gates and qubits), intended for developers to use within Jupyter notebook or equivalent.
QC toolkits also function as access points to physical QPUs and simulators on cloud, which reinforces this usage pattern.
Furthermore, QC developers' interests can change; for example, Jeng's primary research focus was QML (application-side), yet she became interested in pulse-level optimization because QML requires lots of operations which need breakthrough optimization techniques.
Thus, cascading different levels of detail within the same interface and allowing developers to find them whenever they want makes more sense than forcing developers to learn different sets of tools.

\bpstart{(P3) Apply usability standards to QC tools}
During our design iterations, replication, sharing, and context switch were recurring, intertwined topics.
Smith and her collaborators had to come up with custom methods to share circuits and outcomes in a reliable way, and Kim has experienced too basic documentations lacking examples and detailed descriptions.
In this case, applying widely known usability techniques can empower QC researchers and practitioners.
For instance, a common method for exporting and retrieving experiment setups for running circuits (\eg~benchmark codes) and the results (\eg~measurement outcomes) can better facilitate collaboration among QC researchers and practitioners.
Given that understanding QC theory, scripting in Python, and comprehending dense code bases are different skills, QC tools should enable navigating APIs for different tasks, for example, on an overview dashboard.
In doing so, a usable QC interface will need to minimize context switching by offering on-demand information at where developers are working because frequent context switching and the lack of rich documentation can negatively affect productivity in programming~\cite{myer2014,myers2016,shrestha2020}.

\begin{figure*}[!t]
    \centering
    \includegraphics[width=\textwidth]{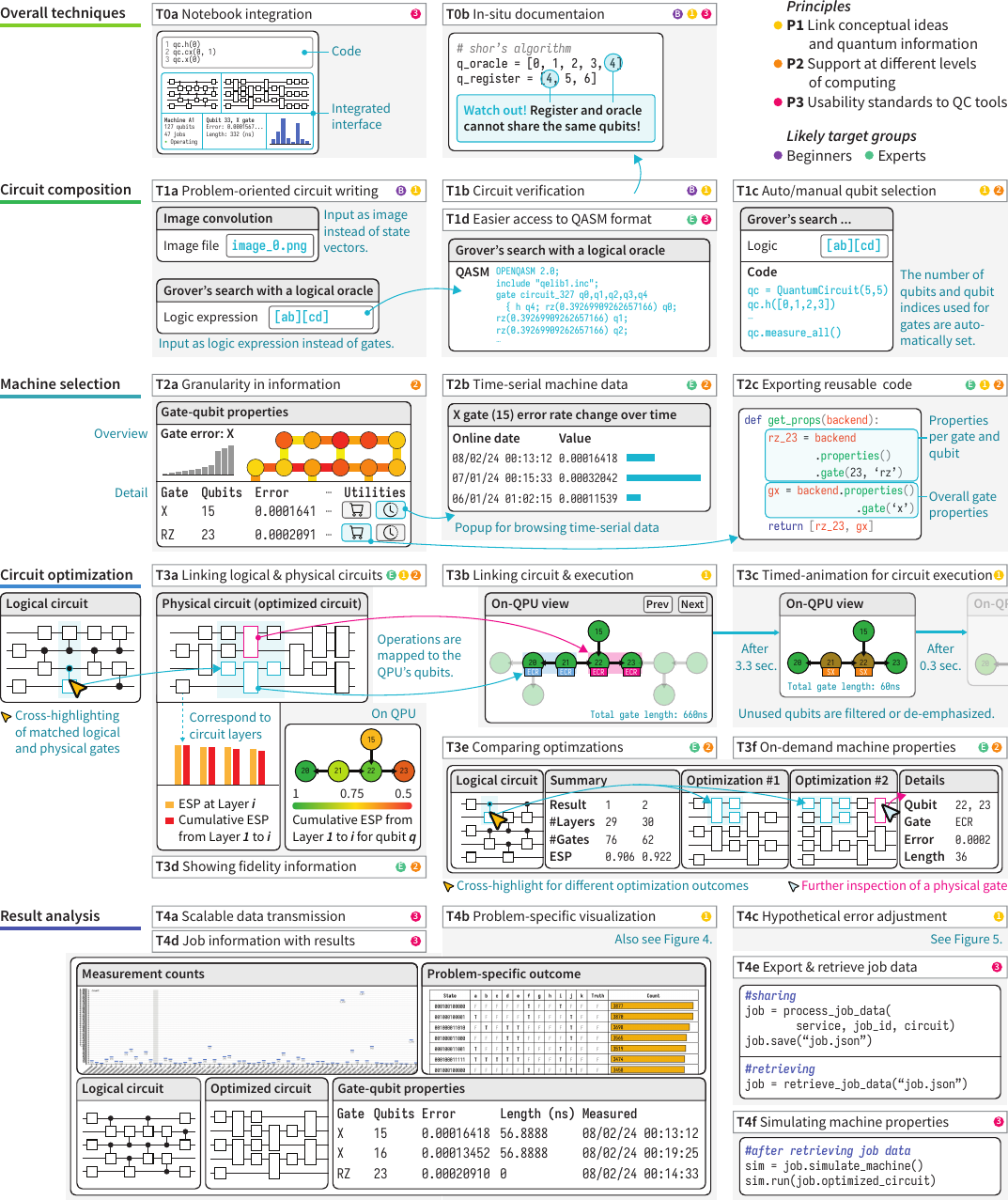}
    \caption{An overview of our interaction techniques for usable QC interface.}
    \label{fig:techniques}
    \Description{QC interaction technique overview in terms of tasks and design principles.}
\end{figure*}

\section{Interface Techniques for Usable QC}\label{sec:techniques}

\noindent
Based on our principles, we propose interaction techniques to guideline usable QC interfaces, implemented as high-fidelity prototypes.
As overviewed in \autoref{fig:techniques}, we introduce individual techniques in terms of high-level tasks: circuit composition, machine selection, circuit review and optimization, and result analysis.
These techniques target those who are learning or practicing QC development skills, compared to those who are learning more theoretical side of QC.
After sketching high-level ideas of those techniques below, we present their use cases in scenarios in the next section.
We provide further technical details in Supplementary Material including the source codes and notebooks.

\subsection{Overall Interface Techniques}\label{sec:techniques:overall}
We first propose two overall techniques: integration with computational notebooks and in-situ documentation.

\bpstart{(T0a) Integration with computational notebooks}
As noted earlier, tools and libraries for QC offer APIs to be used in computational notebooks.
While prior work and existing tools mainly employ stand-alone interfaces, it is cumbersome to export and import data from those interfaces to computational notebooks (\eg~matching file paths, restructuring data).
Thus, our prototypes integrate features for different tasks within the same working environment (\textbf{P3: Baseline usability support}), which we decided as computational notebooks given its wide adoption by current QC developers~\cite{ashktorab2019:hqci}.

\bstartnc{(T0b) In-situ documentation} offers relevant explanations within interfaces, obviating the needs for exploring documentations (\textbf{P3: Baseline usability support}).
In-situ documentations can also encourage QC developers with limited prior knowledge to engage with technical information by making them more understandable (\textbf{P1: Link conceptual and physical}). 
For instance, beginners may not be able to understand some terms like `T1' and `T2'.
Instead, our interface notes that `The T1 and T2 of a qubit stand for how long the qubit can hold its state and phase, respectively. After this time, it is hard to guarantee that the qubit is holding its desired state and phase.' in the interface for exploring machine information.

\subsection{Techniques for Circuit Composition}\label{sec:techniques:write}

We introduce the following techniques to support composing a gate-based QC circuit.

\bpstart{(T1a) Problem-oriented circuit writing}
When using most QC tools, developers have to specify a circuit using atomic gates like H, X, and CNOT gates except some subroutines that are mathematically well generalized, such as quantum Fourier transformation (QFT)~\cite{coppersmith2002:qft}. 
While graphical circuit composers~\cite{ibmComposer,quirk} enables visual programming of QC circuits, they are limited to reduce this inherent gap between conceptual ideas and low-level circuit designs.
Developers still have to specify low-level gates, and it is less scalable for circuits with more than a few qubits~\cite{ashktorab2019:hqci}.
When adopting templates for well-known algorithms, developers still need to update those templates, fix needy-greedy details, and deal with unexpected downstream effects.
Instead, our prototype interface for circuit composition allows for choosing a problem to encode by incorporating prior methods individually proposed by QC researchers (\textbf{P1: Link concepts and quantum}), such as truth tables~\cite{revkit}, image convolution~\cite{jeng2023:convolution}.
Once choosing a problem to solve, then developers can provide parameters using interpretable values like integers, image files, logical expressions, \etc

\bpstart{(T1b) Circuit verification}
Writing a QC circuit involves technical constraints like setting the correct number of qubits to use and updating operation parameters.
When editing circuits, for example, developers may fail to checking those constraints, which commonly happens in code recycling~\cite{mili1995:softwarereuse}.
In the best case, developers just need to debug them, whereas wrong results, the worst case, can cause a huge financial cost for redoing everything.
Thus, our interface verifies a circuit while writing it, so that QC developers can flexibly write and revise their circuits with reduced concerns in technical constraints (\textbf{P1: Link concept and quantum}).

\bpstart{(T1c) Auto/manual qubit selection}
Many QC tools require setting the number of qubits in a circuit prior to adding gates, which is counter-intuitive given that dynamic memory allocation is common in modern programming languages.
The automated qubit selection in our interface lets developers specify a problem and then automatically figures out which qubits (for operations) and classical bits (for measurements) to use to encode that problem (\textbf{P1: Link concept and quantum}).
For example, when an image file is uploaded, this method automatically computes the number of qubits and selects the qubit indices to encode that image file.
To ensure the flexibility in circuit composition, our interface allows for toggling off this automated qubit selection (\textbf{P2: Support at different levels}).

\bpstart{(T1d) Easier access to QASM format}
While QC developers use Open Quantum Assembly Language (or QASM)~\cite{qasm} as a common circuit representation method across different QC libraries, it is difficult to use them.
As basic usability support (\textbf{P3}), our interface allows for saving or copying QASM expressions for circuits written interactively along with Python codes.

\subsection{Techniques for Machine Selection}\label{sec:techniques:macine}

When choosing a QPU, a developer needs to inspect available QPUs, which currently remains tedious.
Below, we introduce related techniques implemented in a dashboard-like interface for a Jupyter notebook setting.

\bpstart{(T2a) Granularity in information}
QC developers need QPU data with varying levels of detail, depending on their goals and expertise. 
For instance, error rate distributions would suffice for beginners while experts would need individual error rates.
Thus, our QPU data dashboard provides overall statistics in visualizations and allows for browsing details in tables and widgets (\textbf{P2: Support at different scales}).
This strategy propagates to the below techniques.

\bpstart{(T2b) Time-serial machine data}
For debugging purposes, QC developers often need to inspect the physical properties of a QPU that change over time.
For example, developers may choose to run the same circuit in batches because a single run can only have up to a certain number of shots.
Then, they need to check the physical properties like gate errors at the time of each batch.
While machine providers like IBM and AWS regularly update relevant information, accessing and navigating time-serial information (\eg~using visualizations) of different properties can be cumbersome because developers have to manually change multiple parameters with current APIs and visualize them to see the trend.
Instead, our QPU data interface allows for optionally selecting reference time points for the properties (\textbf{P2: Support at different scales} \& \textbf{P3: Baseline usability support}). 
Then, the dashboard shows a button for each property by clicking which a small widget shows time-serial visualization for that property.

\bpstart{(T2c) Exporting reusable code}
The online platforms of QPU providers offer visualizations for machine properties, but developers need to make API calls to use them for their work, which is a tedious task as discussed earlier.
Thus, our QPU data dashboard generates a reusable code snippet for selected properties of a QPU (\textbf{P3: Baseline usability support}).
A code snippet takes a function format to ensure its reusability for different QPUs.

\subsection{Techniques for Circuit Review and Optimization}\label{sec:techniques:view}

A QC developer needs to optimize a circuit before executing it.
The optimization quality depends on circuit structure, QPU properties, optimization algorithms, and random seeds.
Typically, circuits become a lot longer (two to ten times) after optimization, so being able to visually and concisely inspect optimization results is necessary. 

\bpstart{(T3a) Linking logical and physical circuits}
QC developers who are testing or learning optimization strategies need to be able to inspect details (\textbf{P2: Support at different scales}).
Thus, our interface juxtaposes a logical circuit and its optimized physical circuits. 
Furthermore, to support directly comparing logical and physical circuits (\textbf{P1: Link concept and quantum}), the interface enables cross-highlighting of corresponding logical and physical gates on a circuit diagram.
We note that there is no deterministic way to map logical and physical circuits, and the most exhaustive method (brute-force) that compares every $x$-consecutive gates has factorial (!) time complexity while not guaranteeing a perfect match.
Thus, we applied a greedy approach that compares the first $k$ gates to the first $k-1$ gates, reducing the time complexity to a polynomial time. 
Our Supplementary Material further details this.

\bpstart{(T3b) Linking a physical circuit to execution}
Circuit optimization essentially places specified operations on physical qubits.
The locations of physical qubits convey important properties like distances between qubits; for example, the more distant two qubits are, the more Swap gates a circuit needs, inducing more error.
However, traditional circuit diagrams exclude such location information.
Thus, our prototype shows selected qubits on a QPU chip and gates assigned those qubits to better assist understanding how circuits operate on a machine (\textbf{P1: Link concept and quantum}).

\bpstart{(T3c) Timed animation for circuit execution}
The prior approach, however, can only show gates in the same layer at a moment, which needs a way to show transitions between layers, such as animations.
Given the importance of gate operation time, we designed circuit execution animation that is timed by gate operation time (\textbf{P1: Link concept and quantum}), in addition to manual transition. 
Since it is hard to recognize actual gate operation time that ranges within a few hundreds nanoseconds, we made the animation proportionately slower with options to adjust the speed.

\bpstart{(T3d) Showing fidelity information}
Mapping gates on a machine diagram makes it more explicit to display how the fidelity of a circuit on a QPU decays over time. 
Fidelity information means how gate error accumulates on qubits over time.
For instance, it is best if accumulated gate error is well distributed across different qubits rather than concentrated on a few qubits. 
An approach to assessing such fidelity is estimated success probability (ESP)~\cite{nishio2020:esp}, which is defined as the product of the success rates ($1 - \text{error}$) of all the gates involved in a physical circuit.
Thus, we integrated three visualizations within the above techniques:
(1) the ESP value of each layer, (2) the cumulative ESP value at each layer, and (3) the cumulative ESP value of each qubit at each layer.
The ESP value of layer $i$ refers to the ESP value for the gates and qubits in the layer $i$ only.
The cumulative ESP value at layer $i$ is defined as the ESP value for the gates and qubits involved in the first $i$ layers. 
The cumulative ESP value of qubit $j$ at layer $i$ means the ESP value for the gates applied to qubit $j$ for the first $i$ layers.
The first two methods appear along with a traditional circuit diagram (T3a), and the last method is shown on a machine diagram (T3b).
Together with summary ESP values in the interface, these methods allow for inspecting ESP at diverse levels (\textbf{P2: Support at different scales}).

\bpstart{(T3e) Comparing optimizations}
Because different factors of a quantum machine affects optimization outcomes (\ie~physical circuits), it is crucial for more than intermediate-level QC developers to be able to compare different optimization results.
Thus, our circuit optimization interface allows for comparing multiple optimization outcomes of the same logical circuit.
First, a table summarizes all the optimization results in terms of the number of gates, ESP values, \etc (\textbf{P2: Support at different scales}).
Next, the interface allows for browsing detailed information of each optimization outcome like circuit diagrams and detailed fidelity information.
When a gate in the logical circuit is selected, corresponding physical gates are highlighted across physical circuits, which is also summarized in a widget for a concise look.

\bpstart{(T3f) On-demand machine properties}
While mapping between logical and physical circuits will suffice the needs of beginning QC developers, more expert people need to understand further information about physical properties of a machine. 
Thus, our circuit visualizations and on-machine views allows for inspecting individual qubits and gates via on-demand widgets and panels (\textbf{P2: Support at different scales}). 
For example, expert QC developers and researchers can toggle the pulse data of a QPU's laser drives (components that manipulate qubits) along with the on-device view when synthesizing new gates by adjusting pulse.

\subsection{Techniques for Result Analysis}\label{sec:techniques:result}

\noindent 
QC program outputs have its own characteristics, such as bit string-based output formats, a large volume of result data, uncertainty due to machine errors, \etc
To support navigating QC results with these characteristics, we propose the following interaction techniques.

\bpstart{(T4a) Scalable data transmission from machine to interface}
QC program outputs may have a huge volume of data. 
For instance, a 20-qubit circuit results in up to a million different output state vectors with counts, which can be more than 50 MB.
Because computational notebooks typically work on a web browser, 
output data from Python libraries need to be exported in a JSON format and  browser needs to parse it again.
The large volume of output data can cause latency or breakage due to overflows on browser's heap memory.
For instance, we experienced IBM's cloud platform failing to show results for a 15-qubit circuit.
By dividing and streaming output data, our interface can seamlessly be displayed within a computational notebook (\textbf{P3: Baseline usability support}).

\begin{figure}
    \centering
    \includegraphics[width=\linewidth]{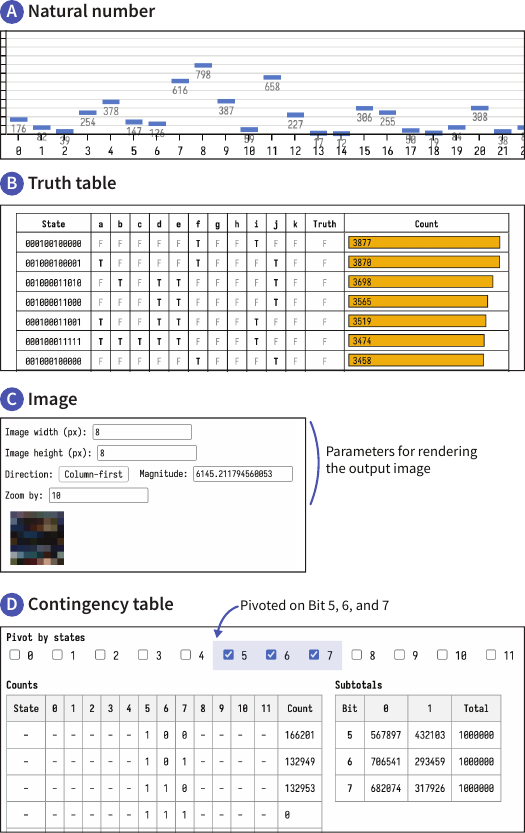}
    \caption{Problem-specific visualizations for (A) natural number, (B) truth table, (C) image, and (D) contingency table.}
    \label{fig:result-vis}
    \Description{Problem-specific outcome visualizations. (A) state vector bit strings are replaced with the corresponding natural numbers in a measurement histogram. (B) State vector bit strings are divided and shown in a truth table view. (C) Measurement outcomes are reformatted as an image, with configuration options. (D) Contingency table for categorical data exploration which can be pivoted by qubits.}
\end{figure}

\bpstart{(T4b) Problem-specific visualization}
QC program outputs are essentially the count of the bit string of each state vector measured.
QC developers tend to need to reformat those bit strings into more interpretable forms like natural numbers, categories, \etc, which requires profound knowledge---yet, many QC tools do not offer relevant methods. 
Thus, our interface offers several methods for visualizing QC program outputs in problem-specific formats (\textbf{P1: Link concept and quantum}), such as integer-based histogram, image, and truth table, with interactions for browsing details.

\bpstart{(T4c) Hypothetical error adjustment}
Given that QC program outputs include some noises occurring due to gate errors, the counts of these bit strings have uncertainty to a degree.
Yet, checking where gate errors actually happened is not realistic and practically impossible because they happen within a few nanoseconds. 
When uncertainty is not easily tractable, a simulation-based method like Monte-Carlo approximation~\cite{mooney1997:monte} is widely used in statistics~\cite{christos2001:monteCarlo}.
Thus, we designed a hypothetical error adjustment method for approximating gate errors using Monte-Carlo simulation to directly present noise from a machine in a more interpretable form (\textbf{P1: Link concept and quantum}) and hence to provide better contexts for collaborative QC programming like benchmarking (\textbf{P3: Baseline usability support}).
We provide detailed computation in Supplementary Material.

\begin{figure}
    \centering
    \includegraphics[width=\linewidth]{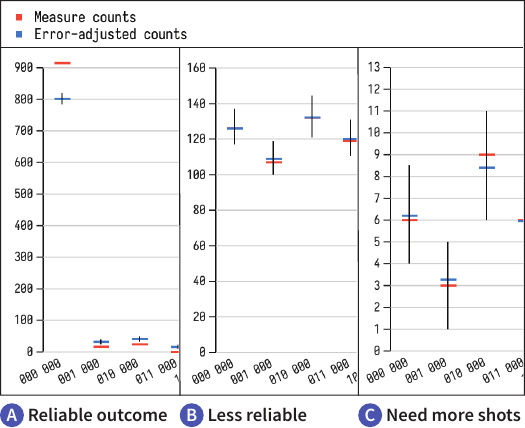}
    \caption{The hypothetical error adjustment technique shows when counts are reliable (A), less reliable (B), and in need of more shots (C).}
    \label{fig:result-hea}
    \Description{Hypothetical error adjustment visualization example, comparing outcomes for reliable, less reliable, and nearly random measurements.}
\end{figure}

This simulation results in an uncertainty visualization like \autoref{fig:result-hea}.
The orange ticks represent the measured count and the blue ticks indicate the mean of hypothetical error-adjusted counts.
The black line behind a blue tick shows the 95\% confidence interval of the corresponding hypothetical error-adjusted count.
Note that this procedure has a general tendency to level the count values toward the center (\ie~the number of shots divided by the number of possible state vectors) because higher state vector counts will decrease whereas lower counts will increase. 
This visualization can help QC developers to assess whether they have ran the circuit with sufficient shots to differentiate state vector counts and if their circuit is returning what they have intended. 
For instance, \autoref{fig:result-hea}A shows a well-differentiated outcome while \autoref{fig:result-hea}B and C indicate the needs for updating a circuit or increasing the number of shots to obtain a clear outcome.

\bpstart{(T4d) Circuit and machine information with results}
Developers would often want to inspect their QC program results with physical properties of a machine as well as other relevant information like the original circuit.
We remind that developers often get outcomes 1-2 days after they submit a circuit to a QPU.
However, current tooling only supports inspecting outcomes, machine properties, and circuit designs in a separate way.
Thus, our result analysis dashboard includes information about the machine and circuit for the results (\textbf{P3: Baseline usability support}).

\bpstart{(T4e) Export and retrieve job data}
A \textit{job} is a common term to refer to the execution and results data of a circuit.
Sharing job data is highly important for collaborative QC development and benchmark. 
To share job data, a developer needs to manually combine the logical and physical circuit designs in a QASM format, state vector counts, and physical properties of a QPU used for the circuit.
Thus, we developed APIs that combine those into a single data object; export into a shareable file; and retrieve the data from that file (\textbf{P3: Baseline usability support}).
For example, once loaded, the job data object provides methods that easily visualize the QC outcomes, circuits, and machine properties (\eg~opening an interface for result analysis).

\bpstart{(T4f) Simulating machine properties}
Given the high cost of running actual QPUs, QC researchers and learners use a simulator for replication.
The replication could take place a while after the initial execution, so the machine properties might have changed already.
Given that the previous technique stores the relevant machine properties (\eg~QPU chip design, error rates), we developed an API that automatically produces a QC simulator that shares the same machine properties with those of the original QPU in order to make it easier to replicate and benchmark QC programs (\textbf{P3}).

\subsection{Implementation Details}
We implemented the above interaction techniques as high-fidelity prototypes.
We implemented a \textit{circuit writer} for circuit composition techniques, a \textit{machine explorer} for machine selection techniques, a \textit{circuit viewer} for circuit review and optimization techniques, and a \textit{result viewer} for result analysis techniques.
We built these interfaces using Svelte.js~\cite{svelte} and embedded them in Jupyter Notebook~\cite{jupyterNotebooks} using AnyWidget~\cite{anywidget}.
These interfaces are available as open-sourced libraries with demo notebooks\footnote{\url{https://see-mike-out.github.io/patoka} or \texttt{pip install patoka}.}. 
\section{Use Cases}\label{sec:demo}

\noindent We demonstrate the above interaction techniques for usable QC interfaces in use case scenarios representing our prior QC usages and challenges.
In particular, we show three scenarios: (1) how the circuit writer and result viewer assist a beginner (James\footnote{Pseudonyms.}) with learning Shor's algorithm; (2) how those tools help a QC practitioner (Paula) with running a QML program; and (3) how the machine explorer and circuit viewer supports a QC researcher (Laura) with examining circuit optimization strategies.

\begin{figure}
    \centering
    \includegraphics[width=\linewidth]{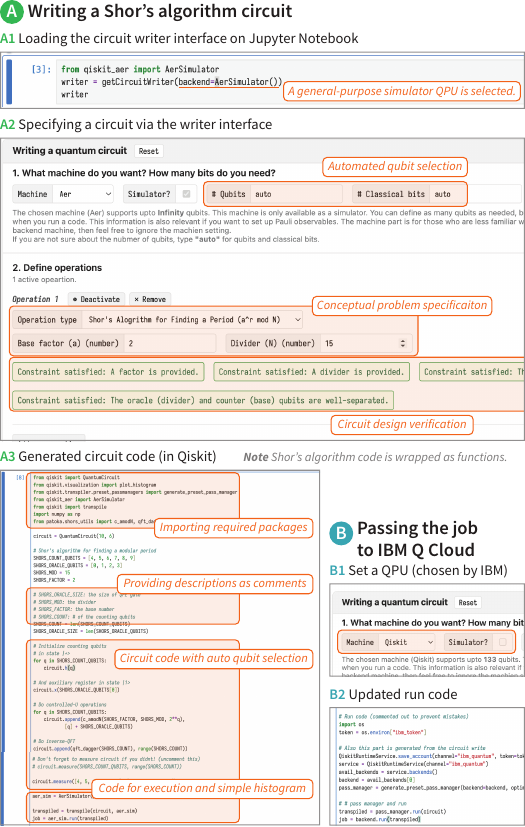}
    \caption{A case study for learning Shor's algorithm (\autoref{sec:demo:case1})---Part 1. (A) Composing a QC circuit for Shor's algorithm using an interactive circuit writer that provides automated qubit selection and circuit verification. (B) Modifying the circuit to choose an available QPU from IBM Q Cloud, and submitting the job.}
    \label{fig:case-1a}
    \Description{The first case study for learning Shor's algorithm, part 1. (A) Writing a Shor's algorithm circuit. A1. Loading the circuit writer interface on Jupyter Notebook. A2. Specifying a circuit via the writer interface. A3. Generated circuit code in Qiskit. (B) Passing the job to IBM Q Cloud. B1. Set a real QPU, chosen by IBM. B2. Updated run code for a real QPU.}
\end{figure}

\begin{figure}
    \centering
    \includegraphics[width=\linewidth]{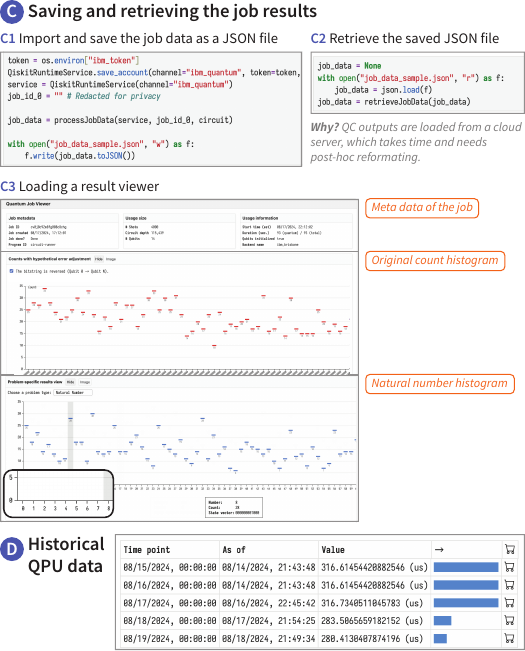}
    \caption{A case study for learning Shor's algorithm (\autoref{sec:demo:case1})---Part 2. (C) Saving and retrieving the job data, and checking the measured outcome. (D) Check the historical QPU data.}
    \label{fig:case-1b}
    \Description{The first case study for learning Shor's algorithm, part 2. (C). Saving and retrieving the job results. C1. Python codes for importing and saving the job data as a JSON file. C2. Python codes for retrieving the saved JSON file. C3. Loading a result viewer on a Jupyter Notebook. (D) Retrieving historical QPU data with a visualization.}
\end{figure}

\subsection{Case 1: Learning Shor's Algorithm}\label{sec:demo:case1}

\noindent 
To apply what he has learned from QC courses, James is trying to run a Shor's algorithm circuit, one that divides an integer into factors, on a real machine.
However, trying out different numbers for Shor's algorithm is tricky and requires more profound knowledge than what typical beginners would have.
Hence, James is going to use an interactive circuit writer on a Jupyter notebook.

First, James imports an circuit writer for an \textit{AerSimulator}, a sandbox simulator with an unlimited number of qubits offered by Qiskit (\autoref{fig:case-1a}A1). 
In the interface, James sets \textbf{automated qubit selection (T1c)} given that he is not yet aware of how many qubits the circuit will need.
After selecting `Shor's algorithm' subroutine, James sets a base factor to 7 and a divider to 15 (\textbf{T1a: Problem-oriented circuit writing}).
As James updates the inputs for this subroutine, the interface \textbf{verifies the circuit (T1b)} and tells whether James has provided required inputs and they are coherent.
With confidence, thus, James extracts the actual Qiskit code from the interface (\autoref{fig:case-1a}A1), which he can paste into the next cell in the current notebook. 
The code includes all the required packages, comments for descriptions, circuit design, and optimization and execution functions.

After testing this code on the simulator, James now wants to pass the circuit to an actual QPU.
Aware that IBM Q Cloud offers free 10-minute credit per month, James sets the device to `Qiskit' in the circuit writer and turns off the simulator (\autoref{fig:case-1a}B1), which results in an updated run code for an actual machine (\autoref{fig:case-1a}B2).
By running this code (simply hitting Ctrl + Enter keys), James submits the transpiled circuit to IBM Q Cloud. 

Running a circuit on a cloud QPU and obtaining the job data takes some time (up to a few days).
Plus, reformatting the loaded data from the cloud is cumbersome.
Thus, James calls a function (\texttt{processJobData} in \autoref{fig:case-1a}C1) to save the job once it is done, and retrieves it (\texttt{retrieveJobData} in \autoref{fig:case-1a}C2; \textbf{T5e}).
Then, James loads a result viewer that includes meta data, traditional histogram, and natural number-based histogram.
The meta information concisely tells him whether the circuit was run successfully (\textbf{T5d}) without looking for codes to retrieve the data (\textbf{T0a}, \textbf{T0b}).
The histogram with natural numbers (\autoref{fig:case-1a}C3; \textbf{T5b: problem-specific visualization}) helps James to easily figure out the potential answers, $7^{8}$ (mod $15 = 1$), $7^{12}$ (mod $15 = 1$), and so on.
Given the large output data size (39.4MB), Qiskit's online platform on the browser failed to show the job outcome, whereas this interface can smoothly show it (\textbf{T5a: Scalable data transmission}). 
Lastly, to make sure that the machine was in a good status, James opens a machine explorer by passing a list of dates after and before the job. 
The machine explorer provides a pop-up to show the \textbf{historical QPU data} (\textbf{T2b}), as illustrated in \autoref{fig:case-1b}D.

\begin{figure}
    \centering
    \includegraphics[width=\linewidth]{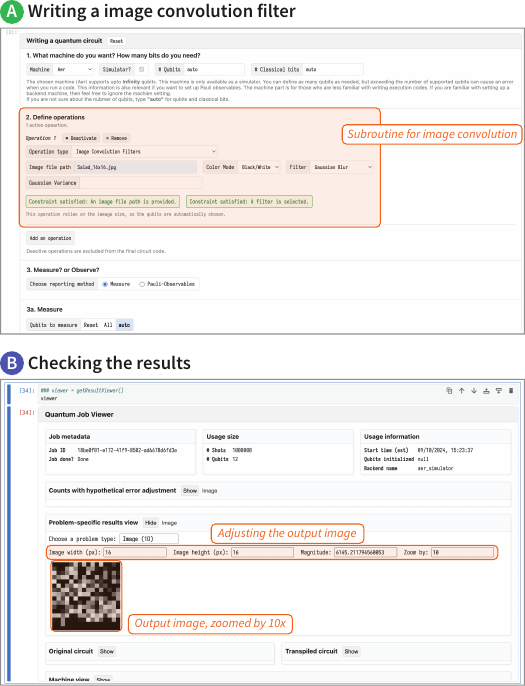}
    \caption{A case study for running an image convolution filter for quantum machine learning (\autoref{sec:demo:case3}). (A) Writing a image convolution circuit. (B) Checking the outcome image.}
    \label{fig:case-3}
    \Description{The second case study for quantum machine learning. (A) Writing a image convolution circuit. (B). Checking the outcome image with interactive configuration options.}
\end{figure}

\subsection{Case 2: Running Quantum Machine Learning}\label{sec:demo:case2}
An important topic in QML is developing subroutines, or \textit{ansatzs}, that can be easily chained to achieve a larger ML network~\cite{ansatz}.
Paula is currently testing which filter to apply for her QML project.
To do so, Paula opens an interactive circuit writer on her Jupyter notebook (\autoref{fig:case-3}A).
In this interface, Paula can provide the image file, the image color dimension, and filter type (\textbf{T1a: Problem-specific circuit writing}).
This generates a code snippet for running this circuit, which saves few hundred lines of code if she did manually.

Once running the circuit on a simulator, Paula saves the resulting job data as a JSON file and retrieves it (\textbf{T5e}).
Then, Paula loads a result viewer to check how the filter worked.
The problem-specific visualization (\textbf{T5b}) converts the measurement outcomes (the bit string and count of each state vector) to an image (\autoref{fig:case-3}B).
Given the image is small, the interface also offers zooming it.

\begin{figure}
    \centering
    \includegraphics[width=\linewidth]{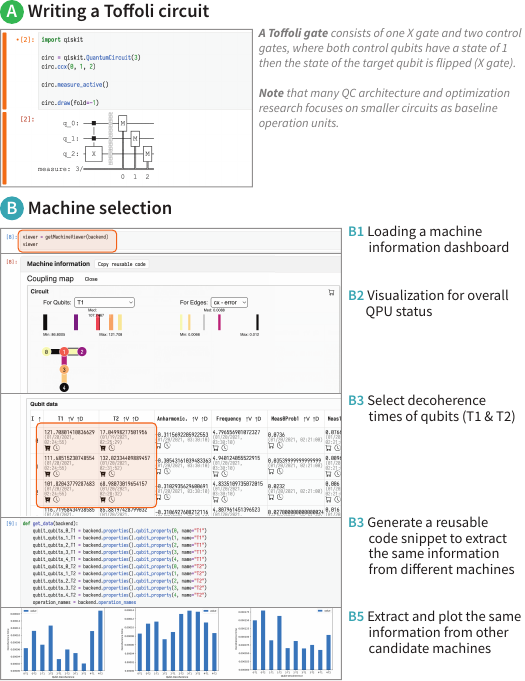}
    \caption{A case study for circuit optimization (\autoref{sec:demo:case2})---Part 1. (A) Writing a Toffoli gate. (B) Exploring the machine date with a reusable code snippet.}
    \label{fig:case-2-1}
    \Description{The third case study for circuit optimization, part 1. (A) Manually writing a Toffoli circuit. (B) Machine explorer interface for machine selection. B1. Loading a machine information dashboard. B2. Select decoherence times of qubits (T1 and T2). B3. Select basis gates. B4. Generate a reusable code snippet to extract the same information from different machines. B5. Extract the same information from other candidate machines.}
\end{figure}

\begin{figure}
    \centering
    \includegraphics[width=\linewidth]{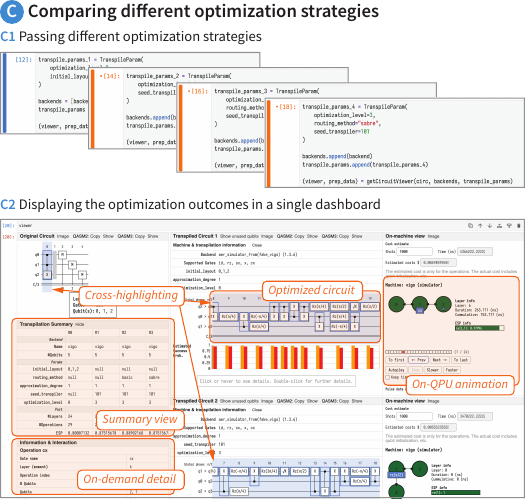}
    \caption{A case study for circuit optimization (\autoref{sec:demo:case2})---Part 2. (C) Updating optimization strategies and comparing the outcomes.}
    \label{fig:case-2-2}
    \Description{The third case study for circuit optimization, part 2. (C) Comparing different optimization strategies. C1. Passing different optimization strategies. C3. Displaying the optimization outcomes in a single dashboard with a summary table and cross-highlighting interaction for the differences in them.}
\end{figure}

\begin{figure}
    \centering
    \includegraphics[width=\linewidth]{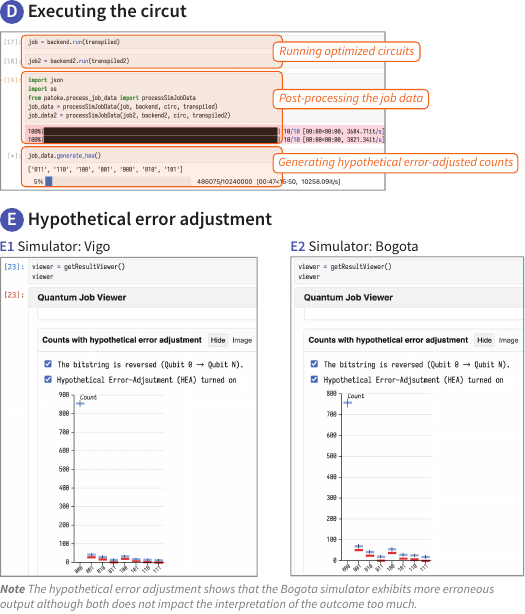}
    \caption{A case study for circuit optimization (\autoref{sec:demo:case2})---Part 3. (D) Running a fully optimized circuit with the basic routing method on \texttt{Vigo} and \texttt{Bogota} simulators. (D) Checking hypothetical error adjustment for the measurement counts.}
    \label{fig:case-2-3}
    \Description{The third case study for circuit optimization, part 3. (D) executing the circuit. Once, the circuit is run on the specified simulators, the job data objects are post-processed and hypothetically error-adjusted measurement outcomes are generated. (E) Hypothetical error adjustment visualization. E1. Vigo simulator. E2. Bogota simulator. Note The hypothetical error adjustment shows that the Bogota simulator exhibits more erroneous output although both does not impact the interpretation of the outcome too much.}
\end{figure}

\subsection{Case 3: Studying Circuit Optimization Strategies}\label{sec:demo:case3}
\noindent
Because each QPU displays a unique noise profile (\eg~gate error, decoherence time), circuit optimization is necessary to maximize the quality of the output.
Yet, optimization is challenging because it must consider an exponentially-growing number of possibilities.
To navigate this space, Laura wants to compare how operating a Toffoli gate may result in different outcomes depending on optimization strategies given the same QPU\footnote{This is a highly simplified version of what Laura would typically do for the demonstration purposes.}.
Simpliy put, a Toffoli gate can be understood as a C-NOT gate that has an additional control qubit (\ie~a 3-qubit gate). 
Toffoli gates frequently appear in QC applications like arithmetic circuits, oracles, and reversible logic.
To start, Laura first writes a Toffoli gate circuit manually given its simplicity (\autoref{fig:case-2-1}A).

Before testing different optimization methods, Laura wants to pick a simulator to use. 
To do so, Laura first imports a machine explorer interface for a \texttt{Vigo} simulator with five qubits (\autoref{fig:case-2-1}B1).
There are other simulators supported by Qiskit, such as \texttt{Athens}, \texttt{Bogota}, and \texttt{Essex}, exhibiting different properties. 
After overviewing the error rate simulation on the chip visualization (\autoref{fig:case-2-1}B2), Laura recognizes the needs for comparing those values across candidate simulators. To extract those values,
Laura selects decoherence times (`T1' and `T2') from the same dashboard (\autoref{fig:case-2-1}B3).
In this way, Laura can explore machines at varying levels of detail (\textbf{T2a: granularity}).
This selection generates a \textbf{reusable code snippet} for extracting that information (\autoref{fig:case-2-1}B4; \textbf{T2c}).
Using that, Laura extracts and plots the same information from other simulators (\autoref{fig:case-2-1}B5), which can easily be combined as a Pandas data frame.
After inspecting them, Laura decides to use the \texttt{Vigo} simulator.

Laura is interested in four optimization strategies, including a baseline case, because the original research relating to those strategies (Li~\ea~\mbox{\cite{li2019:sabre}}) did not look at the Toffoli gate.
After passing the optimization strategies as Qiskit code (\autoref{fig:case-2-2}C1), Laura imports the circuit viewer (\autoref{fig:case-2-2}C2).
Laura compares duration, gate count, layers and cumulative ESP in the summary table (\textbf{T3d: fidelity information} \& \textbf{T3f: comparing optimizations}), and then checks the actual results via cross-highlighting (\textbf{T3a: linking logical \& physical circuits}).
By doing so, Laura can confirm that the stronger optimization improves the feasibility, showing that this circuit is not a trivial case.

By comparing the metrics of different optimization results within the same interface (\textbf{T3f: comparing optimizations}), Laura learns that
one strategy (\#2) improves fidelity in ESP by about 9\%, compared to the zero optimization (\#0).
The same method shows the best performance in terms of ESP and duration.
While this method have a slightly higher gate count than the other test cases (\#1 \& \#3), it improves ESP slightly and saves the runtime by 360 nanoseconds per shot (which can cause a few hundred dollar difference for medium-scale circuits).

Then, Laura runs the Toffoli circuit on simulators given the small size of the circuit. 
Using the same optimization parameters, Laura runs the circuit on \texttt{Vigo} and \texttt{Bogota} simulators given their common basis gates but with different physical properties (\autoref{fig:case-2-3}D).
After running them, Laura generates hypothetically error-adjusted measurement counts (\textbf{T5c}), which reveals that the \texttt{Bogota} simulator exhibits more erroneous outcome for this case (\autoref{fig:case-2-3}E).
While this amount of error induced by the \texttt{Bogota} simulator does not impact the interpretation of the outcome in this case, Laura expects that this may cause more erroneous outcomes when the circuit size is bigger (\ie~more chances of gate error).

\section{Discussion}\label{sec:discussion}
We designed and prototyped interaction techniques for QC tools and demonstrated them via use cases.
By applying HCI techniques to QC, our work implies that HCI methods can play a canonical role in bridging QC with different topics like programming language, software engineering, and interface design.
From these early steps toward human-quantum computer interaction, we identify the following future directions regarding how HCI can facilitate the adoption of QC for varying tasks.

\subsection{HCI + PL for QC}
\noindent Programming language (PL) research has made programming more robust with type systems, logical with better semantics, and efficient with compilers.
In doing so, HCI and PL can have a synergy as it has shown in classical computing.
As reviewed by Chasins~\ea~\cite{chasins2021:plhci}, by learning from each other, PL grammars can better capture targeted tasks, and user interfaces can provide more compositional and principled interactions.
Currently, we do not write a QC program using some specialized forms (\eg~punch cards) but by building formal expressions based on classical PLs (\eg~Python libraries).
Yet, QC programming expressions currently lack a robust type system and well-defined semantics to encode various problems.
For example, a set of qubits can primarily encode states or phases in a probabilistic way, and their outputs can be interpreted as top-k candidates or as an empirical distribution. 
A program can repeat a subroutine to iteratively update qubit states or to use the information multiple times (so as to control other qubits).
Present QC tools do not offer methods to explicitly indicate those behaviors, but programmers need to come up with other tools like NumPy on their own.

Therefore, a robust PL system for QC program should capture how QC programmers conceptualize such behaviors.
Research in PL for QC can benefit from HCI methods, such as analyzing patterns in programming (\eg~\cite{denny2012:syntaxerror,whalley2021:debugging}), demonstrating them using diverse use cases (\eg~\cite{satyanarayan:vega2016,kim:erie2023}), and operating user studies (\eg~\cite{chen2024:cse,myers2004:npl,myers2016:hcpl}).
Such future work will make our circuit writer and result viewer interfaces operate more robustly and efficiently as well as better match to conceptual ideas.

\subsection{Standardizing Interoperable Representations}
\noindent Consistent data representation methods are also necessary for building interfaces, but missed currently in QC tools.
For example, authentication services like `Sign in via Google' can work via standardized data structures that concretely connect layers of data bases, security services, and interfaces.
While implementing the prototypes, however, we frequently experienced inconsistency in API designs and their outcome formats.
A major effort in our prototyping was formatting QC data in a consistent structure so as to enable on-demand user interactions on data-heavy interfaces like the circuit viewer and machine explorer.
Otherwise, ad-hoc conversion had to occur every time a user clicks an element, which would add up browser-side latency.

Future research in data representation methods can work at the inter-operation of two layers: deeper, storage-efficient methods and shallower, interface-efficient methods.
First, deeper representations can support storing QC data in classical databases, efficiently sharing them among users, and minimizing latency in between-application communication. 
Next, shallow expressions can help interface developers be less concerned with backend-side data structures, reducing the needs for ad-hoc methods.
Lastly, a suite of compilers will play a key role in making those representation methods operate with each other.
For instance, data visualization designs are often expressed as concise design specs (\textit{deep}), which are \textit{compiled} to systematically expressed visual entities (\textit{shallow}) that people see on screen.
Standardizing QC representation methods at varying levels will facilitate the development of useful APIs and collaborative interfaces.

\subsection{Domain-specific QC Tools}
\noindent QC is a specialized computing resource, intending for domain experts to solve particular types of problems.
While we prototyped multiple interfaces that included problem-oriented circuit writing and result visualization, an interesting direction can be one-stop shop interfaces for specific domain areas that select and combine different techniques.
For example, a QC interface for drug discovery and chemistry could help researchers to easily set superposition for molecules, encode chemical reactions as an oracle, and show outcomes in a suitable format (\eg~expressing state vectors as formulae). 
Similarly, simulation problems for physics can offer an editor for physical systems (\eg~a set of ions, photons) and show outcomes as animations for how those systems develop over time.
The scalability of those tools will be an important concern.

Future work in domain-specific QC tools may also need to look at the interplay between graphical and character user interfaces.
While graphical interfaces can reduce the gap between computational artifacts and conceptual ideas, it can be cumbersome to manipulate a lot of object by hand (\eg~dragging a few hundred gates to encode a molecule).
Instead, specifying a problem as abstract formulae can make user interaction easier than graphical manipulation.
Future work could also consider the interplay between QC programming and representation methods.
For instance, QC programming tools can allow for directly typing in chemical formulae, and representations methods can include data about timed evolution of a physical system.

\subsection{Feasible Evaluation Methods}
\noindent While our goal was to propose relatively generic techniques to provide blueprints for next-generation QC tools, each interface in our work would need a designated user study for further evaluation.
Prior work has done user studies on highly narrow tasks (\eg~understanding a circuit~\cite{ruan2024:quantumeyes,wen2024:quantivine}, interpreting quantum states~\cite{ruan2023:venus}, reading error rates~\cite{ruan2023:vacsen}) with \textit{expert users} who have done research in QC.
Yet, future QC interfaces should be able to accommodate those who have limited backgrounds in QC but want to explore the space for potential interdisciplinary work.
Evaluation studies then should be able to address the lack of sufficient QC knowledge in a feasible way; for instance, an hour-long lecture for a single-shot study is unlikely to work.
Example approaches can include a single-shot user study with a well-scoped tutorial (\eg~modularization), observing how communities around tools (\eg~Qiskit user group) change in a long term, and deploying tools as course materials.

\subsection{Limitations}
While our work looks at a wide range of QC tasks, our prototype is mainly built for Qiskit based on its popularity among QC researchers. 
Given that different QC tools exhibit low-level differences, future work is needed to make those interface techniques applicable for other QC tools, including analog quantum computers.
Next, our goal was to propose diverse interaction techniques to show how HCI can support QC practices and hence facilitate in-depth research on individual techniques and tasks. 
While we implemented those techniques as functioning prototypes for a proof of concept and applied those prototypes to realistic use cases with varying levels of expertise, user evaluation-based future research will benefit extending our approaches. 

\section{Conclusion}\label{sec:conclusion}
While expected to provide promising solutions for classically intractable problems, quantum computing (QC) remains challenging for wide adoption across different domains because users need profound technical and physical knowledge to use quantum computers.
To facilitate research and development of usable QC interfaces, we derived a set of high-level principles through design iterations over three months.
Based on these principles, we proposed interaction techniques spanning from circuit composition, to machine selection, circuit optimization, and result analysis.
After implementing these techniques as widgets for Jupyter Notebook, we demonstrate their feasibility via three use cases reflect varying levels of QC expertise from beginner to expert.
We conclude by discussing future interdisciplinary research agenda between HCI and QC.



\bibliographystyle{ACM-Reference-Format}
\bibliography{references}


\end{document}